\documentclass[aps,preprintnumbers, showpacs,notitlepage,twocolumn]{revtex4-1}
\usepackage{graphicx}
\usepackage{epsfig}
\newcommand{\ben}{\begin{eqnarray}}
\newcommand{\een}{\end{eqnarray}}

\begin{document}
\title{Inclusive b-jet production in heavy ion collisions at the LHC }

\date{\today}

\author{Jinrui Huang}
\email{jinruih@lanl.gov} 
\affiliation{Theoretical Division, 
                   Los Alamos National Laboratory, 
                   Los Alamos, NM 87545, USA}

\author{Zhong-Bo Kang}
\email{zkang@lanl.gov} 
\affiliation{Theoretical Division, 
                   Los Alamos National Laboratory, 
                   Los Alamos, NM 87545, USA}

\author{Ivan Vitev}
\email{ivitev@lanl.gov}                   
\affiliation{Theoretical Division, 
                   Los Alamos National Laboratory, 
                   Los Alamos, NM 87545, USA}

\begin{abstract}
Theoretical and experimental advances in understanding light jet production and modification 
in Pb+Pb reactions at $\sqrt{s_{NN}}=2.76$~TeV have been a highlight of the LHC heavy ion program.  
At the same time, the detailed mechanisms of heavy quark propagation and energy loss in dense QCD 
matter are not yet fully understood. With this motivation, we present theoretical predictions for the 
nuclear-induced attenuation of the differential cross section for inclusive b-jet production in 
heavy ion collisions at the LHC for comparison to upcoming data. We find that for transverse 
momenta  $p_T  \gtrsim 75$ GeV both hadronization and mass effects are negligible and  this attenuation 
is comparable to the one observed for light jets. We discuss how the detailed b-jet quenching
patterns can be used to gain new insight into the in-medium heavy flavor dynamics.  
\end{abstract}
\preprint{LA-UR-13-24055}

\maketitle

\section{Introduction}
Collimated jets of hadrons are a dominant feature of high energy particle interactions. 
They have long been  regarded as the main tools for testing the fundamental 
properties of Quantum Chromodynamics (QCD), for 
obtaining information about the structure of the proton, and for searching for new physics beyond the 
Standard  Model~\cite{Campbell:2006wx}. The current 
highest energy hadron collider, the Large Hadron Collider (LHC) at CERN, guarantees an abundant yield of 
large transverse momentum jets~\cite{Aad:2010ad,Chatrchyan:2011qta,Abelev:2013fn}.

High energy jets also provide unprecedented opportunities to study the properties of dense 
QCD matter,  cold nuclear matter (CNM) and 
the quark-gluon plasma (QGP) that is created in  the ultra-relativistic  collisions of 
heavy nuclei. In such reactions, 
large transverse momentum partons traverse a medium composed of colored quasi-particles
and lose energy via induced gluon radiation and elastic 
scattering. This ``jet quenching'' phenomenon  modifies the observables associated  with  
jets relative to the vacuum case~\cite{Vitev:2008rz}. The differences can be used to 
gain insight into the mechanisms of parton interactions in dense QCD matter and to provide 
deeper  understanding of the in-medium parton shower formation. 
For this reason, inclusive and tagged jet production in p+p collisions 
and the corresponding   modification in Pb+Pb reactions   have  been 
intensively investigated by both the ATLAS and the CMS collaborations at the 
LHC~\cite{Aad:2010bu,Chatrchyan:2011sx}.  ALICE collaboration results are also expected
in the near future.  Theoretical predictions and  description of the data  based 
on models of ``jet quenching'' are in good agreement with the experimental 
results~\cite{Neufeld:2010fj,He:2011pd,Qin:2010mn,Wang:2013cia,Young:2011qx}, providing 
valuable information on the {\it light} flavor parton dynamics in the QGP.

Although these successful descriptions are rather encouraging, there are still 
remaining open questions. One well-known difficulty is related to the fact that the {\it heavy} 
quark (charm and bottom) parton-level energy loss
has not been sufficient in the past to explain the observed suppression in the heavy mesons 
(or $e$, $\mu$ coming from their semi-leptonic decays) at both the RHIC and the 
LHC~\cite{Adler:2005xv,ALICE:2012ab}. There are extensive theoretical attempts aiming to 
explain the observed suppression~\cite{Adil:2006ra,Wicks:2005gt,vanHees:2007me,Qin:2009gw,Gossiaux:2008jv}. 
Meanwhile, additional experimental observables are also being  proposed to further test the heavy 
flavor energy loss mechanisms~\cite{Kang:2011rt}. 

In this paper, we take a different approach and present a study of inclusive heavy 
flavor jet production in heavy ion collisions. More specifically, we report first 
results and predictions for the inclusive b-jet cross sections in   $\sqrt{s_{NN}}=2.76$~TeV  
Pb+Pb  collisions  at the LHC. We combine consistently the simulations 
in nucleon-nucleon reactions, which are validated through comparison to the $\sqrt{s} = 7$~TeV 
p+p data at the LHC, with CNM effects and parton energy loss effect 
in the QGP. We study the relation between the attenuation of inclusive 
b-jet production and the physics of heavy quark production and propagation in dense
QCD matter. Predictions on the nuclear modification factor $R_{AA}$ as a function 
of the jet transverse momentum $p_T$ and the jet radius parameter $R$ are given.

The rest of our paper is organized as follows: in Sec.~II we present the evaluation of 
the inclusive b-jet cross section in p+p collisions using Pythia simulations. In Sec.~III 
we describe the simulation of the medium-induced parton shower and the related heavy quark
mass effects. Our phenomenological results for the b-jet production in the Pb+Pb collisions 
at the LHC are given in Sec. IV. We conclude our paper in Sec. V.

\vfill

\section{Inclusive b-jet  production in p+p collisions}
In this section, we present the evaluation of the inclusive b-jet cross section in 
p+p collisions using  Pythia~8~\cite{Sjostrand:2007gs}. Specifically, the 
Pythia~8 event generator with the CTEQ6L1 parton distribution functions~\cite{Pumplin:2002vw} 
is employed to simulate the inclusive jet events in the p+p collisions for center-of-mass 
energies  $\sqrt{s}=7$~TeV and $\sqrt{s} = 2.76$~TeV. This generator utilizes leading-order 
perturbative QCD matrix element for $2\to2$ processes, combined with a leading-logarithmic 
$p_T$-ordered parton shower, and the Lund string model for hadronization. Furthermore, 
the SlowJet program is used for the jet clustering with an anti-$k_T$ algorithm~\cite{Cacciari:2008gp}, 
which has been checked to yield the same results as FastJet~\cite{Cacciari:2011ma}. 
The b-jets are defined to be jets that contain at least one b-quark (or ${\rm \bar b}$-quark) 
inside the jet cone. Thus, to further identify  the b-jets from the inclusive jets 
found by the SlowJet, a b-quark (or ${\rm \bar b}$-quark) 
is assigned to a jet if the radial separation from the reconstructed jet axis is  $\Delta R < R$. 
$R$ is the jet radius parameter and 
$\Delta R = \sqrt{\left(\Delta\phi\right)^2+\left(\Delta y\right)^2}$, 
where $\phi$ and $y$ are the azimuthal angle and the rapidity, respectively. 

\begin{figure}[!t]
\psfig{file=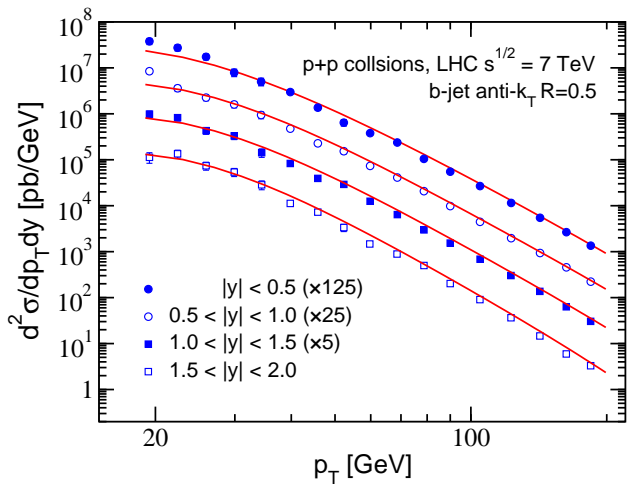, angle=270, width=0.95\linewidth}
\caption{Differential cross sections versus $p_T$ are shown for inclusive b-jets in four 
different rapidity $y$ regions at the LHC, $\sqrt{s}=7$~TeV p+p collisions. Red lines are from 
the Pythia~8 simulations, using the anti-$k_T$ jet algorithm with $R$ = 0.5.  Circles, open-circles, 
squares and  open-squares correspond to the CMS experimental results~\cite{Chatrchyan:2012dk}.}
\label{b-cross}
\end{figure}

In Fig.~\ref{b-cross}, we evaluate the inclusive b-jet differential cross section $d^2\sigma/dydp_T$ 
with $R=0.5$ in p+p collisions at $\sqrt{s}$ = 7~TeV as a function of jet transverse momentum $p_T$ 
at different rapidity regions using Pythia~8 (red lines). We compare the simulations 
to  recent CMS experimental data (circle, open-circle, square and open-square 
data points)~\cite{Chatrchyan:2012dk}. We find that for all four rapidity regions the Pythia 
simulations agree well with the experimental data. Note, that we have included hadronization effects 
in this comparison. Let us elaborate further on such effects. Since all interactions in dense QCD matter that we consider in this paper happen at the partonic level, it is important to assess how  hadronization 
affects the b-jet cross section. In Pythia simulations one can easily turn off hadronizaiton effects, but 
still keep the parton shower and obtain the b-jet cross section from the clustering of the 
 final-state partons. We find that, for all  jet radii relevant to our study ($0.2\leq R\leq 0.7$), 
noticeable effects  of hadronization exist only at relatively low transverse momentum  
($p_T\lesssim 30$ GeV) and small jet radius parameter $R=0.2, 0.3$. This is consistent with the 
basic premises of jet physics, namely, that it accurately reflects the parton level QCD dynamics.
Since we are mainly interested  in the  high $p_T$ region, the hadronization effects are 
negligible in our study of nuclear modification  of the b-jet production. 

In the following sections we will present the nuclear modification of b-jet production in 
Pb+Pb collisions at $\sqrt{s_{NN}}=2.76$~TeV. In order to study  energy loss effects 
and the medium-induced parton shower, we need  detailed information on the flavor content of 
the b-jets produced in the p+p collisions at $\sqrt{s}=2.76$~TeV. Heavy quark production 
in hadronic collisions can be generated through various mechanisms~\cite{Norrbin:2000zc} 
that  (in the fixed flavor scheme \cite{scheme})  can be categorized into three classes: 
``heavy quark pair creation'', ``gluon splitting'', 
and all other ``light quark pair creation'' processes. In Pythia simulations, these mechanisms can be easily 
separated and studied through the $2\to 2$ hard partonic scattering, showering, and hadronization. 
In the so-called ``heavy quark pair creation'' process, 
two heavy quarks are produced in the hard subprocess. 
At leading order this is described by $gg\to b\bar b$ and $q\bar q\to b\bar b$. In this case, 
one of the b-quark (${\rm \bar{b}}$-quark) initiates the b-jet. 
In heavy ion collisions, the medium modification of such b-jet behaves more like a 
single b-quark transversing the QGP, and thus has a direct connection to the heavy quark 
energy loss~\cite{Zhang:2003wk}. 
On the other hand, the ``gluon splitting'' mechanism can also generate b-jets, in which heavy quarks 
do not participate in the hard subprocesses, but are produced through the $g\to b\bar b$ channel 
in the parton shower. In Pythia simulation, we collect $gg\to gg$, $qg\to qg$, and $q\bar q\to gg$ 
in this category. The time for such gluon splitting,  $ [p^+,0, {\bf 0}_\perp]_g  \rightarrow  
[ z p^+, ({\bf k}^2 + m_b^2 )/2zp^+ ,  {\bf k}]_b +  [ (1-z) p^+,  
({\bf k}^2 + m_b^2 )/2(1-z)p^+ , -{\bf k}]_{\bar b} $, can be estimated from the inverse
virtuality for this process to occur~\cite{Adil:2006ra}:
\begin{equation}
\tau_{\rm split}=\frac{1}{1+\beta_{b\bar b}} \frac{2z(1-z)p^+}{ {\bf k}^2 + m_b^2 } \;. 
\label{split}
\end{equation}
In Eq.~(\ref{split}) $p^+$ is the large lightcone momentum of the parton, the transverse momentum 
$ |{\bf k}|  \sim \Lambda_{\rm QCD}$ is perpendicular to the direction of the jet, and  
$\beta_{b\bar b}$ is the velocity of the heavy $b\bar b$ pair (we take $m_b = 4.5$~GeV). 
Even for jets of transverse momentum of 200~GeV at midrapidity ($p^+=400$~GeV) the 
splitting time is $\tau_{\rm split} < 1$~fm, much smaller than the typical size of the medium 
$\sim R_{Pb}$.
In heavy ion collisions, the medium modification of these b-jets would resemble that of 
a massive gluon (color octet state) transversing the QGP~\cite{Sharma:2012dy}. Finally, 
all other partonic channels can also generate b-jets through the parton shower. 
These include $gg\to q\bar q$, $qq\to qq$, and $q\bar q\to q\bar q$ in the Pythia simulations. 
Following the same  logic, they  will 
behave like a massive quark (color triplet state) transversing the QGP in the case 
of heavy ion collisions. We denote the fractional contribution to the b-jet cross section 
from the ``heavy quark pair creation'' as $R_b$, the ``gluon splitting'' as $R_g$, and the 
remaining contributions as $R_{\rm other}$. 

\begin{figure}[!t]
\psfig{file=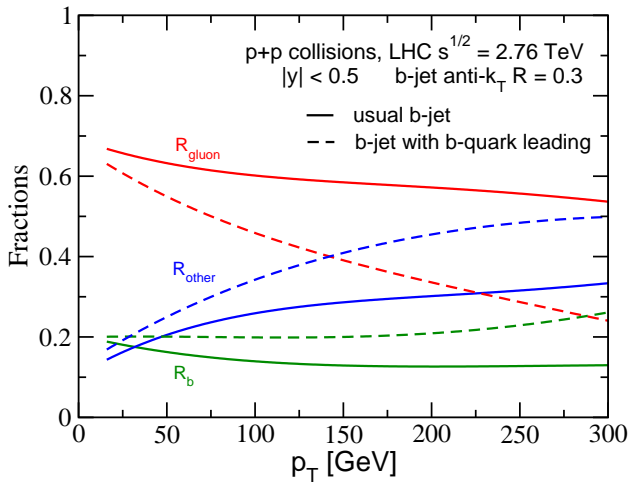, width=0.95\linewidth}
\caption{The relative contribution of different subprocesses  to the inclusive b-jet production 
cross section in p+p collisions at the LHC with $\sqrt{s}=2.76$~TeV. Here,  b-jet production 
is simulated using Pythia~8 with an anti-$k_T$ jet algorithm and $R=0.3$, and we have integrated 
over the central rapidity region $|y|<0.5$. The solid curves are for the commonly defined b-jets, 
while the dashed curves further require that the contained b-quark (or ${\rm \bar b}$-quark) 
is the leading particle in the jet.}
\label{ratio}
\end{figure}

In Fig.~\ref{ratio}, we show these relative contributions as a function of 
the usual b-jet transverse momentum $p_T$ for $\sqrt{s}=2.76$ TeV p+p collisions 
with $R=0.3$ and jet rapidity $|y|<0.5$ (the solid curves). We find that the contribution from 
``gluon splitting'' is dominant in all of the studied $p_T$ range, while the ``heavy quark pair 
creation'' contribution is only around $10\%-20\%$, depending on the $p_T$ range. Our relative 
fractions $R_b$,  $R_g$, and  $R_{\rm other}$  are roughly consistent with previous 
simulations  based on Herwig~\cite{Banfi:2007gu}. Naively, one would like the nuclear 
modification of b-jets to be related to the dense QCD matter effects (energy loss) of a leading b-quark.  
This implies that one has to increase the contribution from ``heavy quark pair creation'', 
as we have explained above. We find that if  we require that the contained b-quark 
(or ${\rm \bar b}$-quark) is also the leading particle inside the jet cone, the fraction $R_b$ can 
increase quite a bit at high $p_T$ and reach $\sim 30\%$ at $p_T\sim 300$ GeV, which is still 
subdominant. To further increase the contribution from the ``heavy quark pair creation'', 
one  realizes the following: for b-jets that originate from heavy flavor creation, only one 
 b-quark  (or ${\rm \bar b}$-quark)   is expected to be inside the same jet cone. On the other hand, for b-jets that 
originate from gluon splitting, the b-${\rm \bar b}$ quark pair is likely collimated 
and, hence, clustered in the same jet. To reduce the relative contribution from ``gluon splitting'', 
one can study  b-jets which contain only a single b-quark inside the jet cone, as has already 
been done by the CDF collaboration at the Tevatron~\cite{Aaltonen:2008de}. Nuclear modification 
of such ``single b-quark jet'' might reflect better the physics of heavy quark energy loss in the QGP.

\section{ Modification of  b-jet production rates in dense QCD matter }

The baseline for the evaluation of the b-jet cross sections in heavy ion collisions are
the corresponding cross sections in the more elementary nucleon-nucleon reactions.
In Ref.~\cite{Ovanesyan:2011xy} it was shown that in QCD the final-state
process-dependent medium-induced radiative corrections factorize in the production 
cross sections for the example of a single jet,
\begin{eqnarray}
d \sigma(\epsilon)^{{\rm jet}}_{\rm quench.} &=&
d\sigma(\epsilon)^{{\rm jet}}_{\rm CNM}  \otimes P(\epsilon)
\,  |J(\epsilon)|  \;. \qquad
\label{fact}
\end{eqnarray} 
In Eq.~(\ref{fact}) $d\sigma(\epsilon)^{{\rm jet}}_{\rm CNM}$ contains only 
cold nuclear matter effects and the differential phase space factors are omitted for brevity.
Furthermore, the functions above depend on the kinematics, such as the jet $p_T$ and $y$, which
we don't write explicitly.   
$P(\epsilon)$ is the probability that the hard parent parton will redistribute 
a fraction of its energy $\epsilon$ into a medium-induced parton shower, and
$\otimes$ denotes an integral convolution.   
$|J(\epsilon)|\equiv |J(\epsilon;R,\omega^{\rm coll})|$ is a phase space Jacobian that further takes into account   
what fraction of the medium-induced parton shower energy is retained inside the jet cone
of radius $R$, as opposed to ``lost'' outside.  The reason that formation of an in-medium parton shower
ultimately leads to a suppression of the jet cross section is two-fold.
First, this shower has a  wider angular distribution relative to 
the vacuum shower. This is true both in the
soft gluon approximation that we use in this paper and for the full ${\cal O}(\alpha_s)$ 
medium-induced  splittings~\cite{Vitev:2005yg}. We recently checked on the example 
of $q\rightarrow qgg$ that the broad angular  
distribution feature of in-medium branchings  holds to  ${\cal O}(\alpha_s^2)$~\cite{Fickinger:2013xwa}.
Second, the partons of the in-medium parton shower can further transfer their energy 
to the strongly-interacting plasma through collisional interactions at a rate 
$ \frac{dE^{\rm coll}}{d\Delta z}  \propto \frac{ \alpha_s^{\rm med} C_{ ( s ) } 
 m_D^2 }{2} \ln \frac{\sqrt{E\, T} }{m_D}$ to leading logarithmic accuracy~\cite{Neufeld:2011yh}.
Here, $m_D = g^{\rm med} T$ is the Debye screening mass for a gluon-dominated plasma of 
temperature $T$ and  
$C_{(s)}$ is the quadratic Casimir for the color representation of the corresponding
state:   $C_{(3)} = C_F = 4/3$, $C_{(8)} = C_A = 3$. 
This energy is also transported away from the direction of the jet through the excitations 
of the medium.

The starting point for the evaluation of the QGP effects in Eq.~(\ref{fact}) 
is the medium-induced gluon distribution, which we calculate to first order in 
opacity~\cite{Vitev:2007ve}:
\ben
k^+ \frac{d^3N^g}{dk^+ d^2 {\bf k}} \!\!  &=&  \!\! \frac{C_{(s)} \alpha_s}{\pi^2} 
 \int  \frac{d \Delta z}{\lambda_{(8)} } 
\int  d^2 {\bf q} \;  \frac{1}{\sigma_{el}}    \frac{d^2\sigma_{el}}{d^2 {\bf q}} \;  
\frac{ 2 {\bf k} \cdot {\bf q} }{{\bf k}^2 ({\bf k}-{\bf q})^2}   
 \nonumber \\
& &    
\times \left( 1 - \cos\left(   \frac{ ({\bf k}-{\bf q})^2}{k^+} \Delta z \right) \right). 
\label{rellm}
\een
In Eq.~(\ref{rellm}), $k^+$ is the large lightcone momentum of the emitted gluon in the
direction of the hard parent parton, ${\bf k}$ is its momentum transverse to the jet axis,
and  ${\bf q}$ is the transverse momentum exchange between the propagating system and  the
QCD medium. $\frac{1}{\sigma_{el}} \frac{d^2\sigma_{el}}{d^2 {\bf q}}$  is the normalized 
differential distribution of the transverse momentum transfers. Note that, as written,
Eq.~(\ref{rellm}) relates to massless partons. Thermal effects in the sub-asymptotic 
temperature regime  are included with an effective mass  ${\bf k}^2 \rightarrow {\bf k}^2 
+ m_D^2 $.  The details of 
evaluating the double differential distribution $k^+ \frac{d^3N^g}{dk^+ d^2 {\bf k}}$
are given elsewhere~\cite{Vitev:2008rz,Vitev:2007ve}. We point out that, in order to 
calculate  the bremsstrahlung  of a collimated multi-parton system that is not 
resolved  by  the momentum exchanges in the medium, we have generalized 
Eq.~(\ref{rellm}) to relate the color state $(s)$ to the representation of the parent 
parton. For example, for $g\rightarrow b\bar b$, $(s)=(8)$ and   
$q \rightarrow q g \rightarrow  q  b\bar b$, $(s)=(3)$.  The mass $M = m_b, 2m_b$ is 
implemented  as follows:  ${\bf k}^2  \rightarrow {\bf k}^2 + x^2 M^2 $, where 
$x = k^+/p^+$  is the lightcone momentum fraction of the gluon  
($k^+ \approx 2 \omega$, $p^+=2E$). An example of such application for a different
observable will be the quenching of the color-octet contribution to quarkonium 
production at high transverse momentum~\cite{Sharma:2012dy}. 

With the QGP-induced distribution of gluons   $\frac{d^3 N^g}{dk^+ d^2 {\bf k}}$ 
and the related $\frac{d^2N^g}{d \omega d r}$ ($\omega$ is the energy and $r$ is the angle)  
of gluons at hand, we can evaluate the fraction $f$ of the  in-medium  parton 
shower energy that is contained in the jet cone of radius $R$:  
\begin{equation}
 f(R,\omega^{\rm coll})_{(s)}= \frac{\int_0^{R} dr
\int_{\omega^{\rm coll}}^E  d\omega \,
\frac{ \omega d^2N^{g}_{(s)}} {d\omega  dr }}
{\int_0^{{R}^{\infty}} dr \int_{0}^E  d\omega \,
\frac{ \omega d^2N^{g}_{(s)}}{d \omega dr} } \; .
\label{fraction}
\end{equation}
In Eq.~(\ref{fraction})   $f(R,0)_{(s)} $ takes into account  medium-induced 
parton splitting effects. On the other hand $f(R^\infty, \omega^{\rm coll})_{(s)}  
= \Delta E^{\rm coll} / E $ is the energy dissipated by the medium-induced 
parton shower into the QGP due to collisional processes.  $\Delta E^{\rm coll}$
is evaluated as in Ref.~\cite{Neufeld:2011yh} and helps solve for $\omega^{\rm coll}$.
Then, for any $R$,   Eq.~(\ref{fraction}) allows to treat the radiative and 
collisional energy loss effects on the same footing.  

Writing down explicitly the phase space Jacobian
$|J(\epsilon)|_{(s)} = 1/\left(1 - [1-f(R,\omega^{\rm coll})_{(s)}]  \epsilon \right)$ and
applying  Eq.~(\ref{fact}) to the case of b-jets we obtain the cross section
per elementary nucleon-nucleon collision: 
\begin{eqnarray}
\frac{1}{\langle  N_{\rm bin}  \rangle}
 \frac{d^2 \sigma_{AA}^{\rm b-jet}(R)}{dydp_{T}} &=& \! \sum_{(s)} \!\!
\int_{0}^1  \!\! d\epsilon \;  
\frac{ P_{(s)}(\epsilon) }{  \left(1 - [1-f(R,\omega^{\rm coll})_{(s)}]  \epsilon \right)  } 
\nonumber \\
&&  \!\!\!\!\!   \times 
 \frac{d^2\sigma^{\rm CNM,LO+PS}_{(s)} \left(|J(\epsilon)|_{(s)} p_{T} \right)} {dy dp_{T}} \; . \qquad
\label{incl}
\end{eqnarray}
Here, the sum runs over the set of final states (s) described in Sec.~II. Their 
relative contribution as a function of $p_T$ was presented in Fig.~\ref{ratio}.
 $d^2\sigma^{\rm CNM,LO+PS} / {dy dp_{T}} $ includes cold nuclear matter effects. Note that,
just like in the QGP,  we consider  physics related to the coherent, elastic and inelastic
parton scattering in large nuclei~\cite{Vitev:2006bi}. Of these effects, at the transverse
momenta and virtualities for the hard parton scattering subprocess that we consider, only
initial-state energy loss may play a role~\cite{Vitev:2007ve}. As we will see in the next 
section, the influence of CNM  effects on b-jet production is relatively small. 

In reactions with heavy nuclei one has to take into account the geometry of the 
collision. The hard parent parton production points are distributed in the transverse
plane according to the  binary collision density $\sim d^2N_{\rm bin}/d^2{\bf x}  $.
In contrast, the gluon-dominated medium is assumed to follow the number of participants
density $ \sim d^2N_{\rm part}/d^2{\bf x} $ and undergoes longitudinal Bjorken
expansion. Specifically, we use an optical Glauber model with an inelastic
scattering cross section $\sigma_{in} = 64$~mb.

\section{Phenomenological results}

We now turn to the phenomenological results for b-jet production in Pb+Pb collisions at the 
LHC. Our simulations are performed at a center-of-mass  energy per nucleon-nucleon pair  
$\sqrt{s_{NN}}=2.76$~TeV for direct comparison to upcoming experimental measurements at the LHC.
Unless otherwise specified, we focus on the most central collisions with average number of 
participants $N_{\rm part} = 360$. As discussed in the previous section, the redistribution of the 
energy of a hard scattered parton in the final state between a vacuum and a medium-induced parton 
shower will lead to a modification of the observed b-jet cross section. It can be quantified 
through the nuclear modification ratio:
\begin{equation}
 R_{AA}^{\text{b-jet}}(p_T; R)  
 =   \frac{  \frac{d^2\sigma^{AA}(p_T ; R  ) }{ dy  dp_T  } }
{ \langle  N_{\rm bin}  \rangle
\frac{d^2 \sigma^{pp}( p_T ; R  ) }{ dy  dp_T  } } \; .
\label{RAAjet} 
\end{equation}
Here, $d^2\sigma^{pp}(p_T ; R)$ and $d^2\sigma^{AA}(p_T ; R  ) $ are the differential cross sections 
in p+p and A+A reactions respectively and $\langle N_{\rm bin} \rangle$ 
is the average number of nucleon-nucleon scatterings for a given centrality class.

\begin{figure}[!t]
\centerline{
\includegraphics[width = 0.95\linewidth]{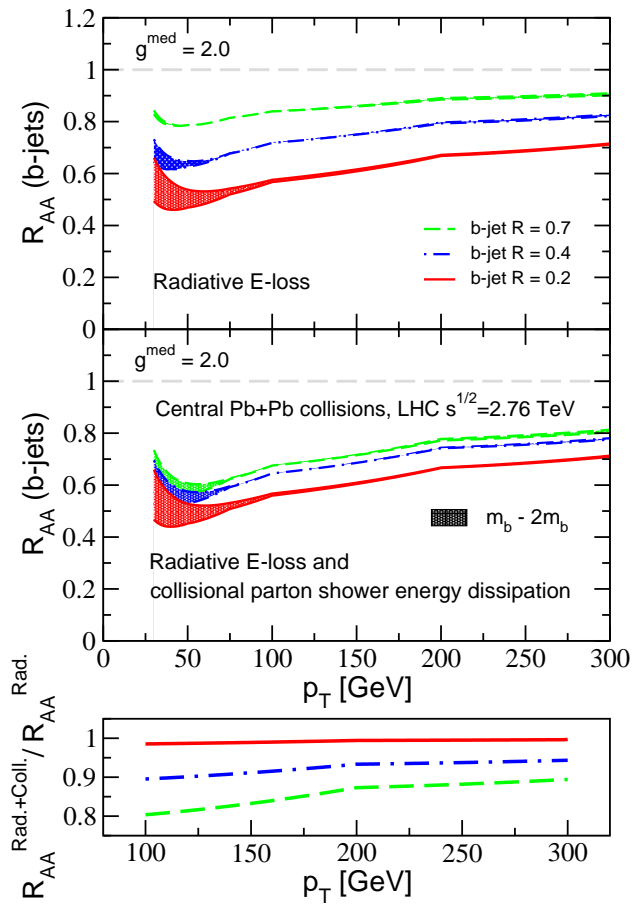}\hskip 0.03\linewidth
}
\caption{The predicted jet radius $R$ dependence of the nuclear suppression for b-jet production
in central Pb+Pb collisions at the LHC at $\sqrt{s_{NN}}=2.76$~TeV is shown verses the jet transverse
momentum. We have chosen radii $R = 0.2,  0.4,  0.7$ and a coupling between the jet and the
medium $g^{\rm{med}}=2$. The upper panel only shows the effect of radiative energy loss 
and the lower panel includes the collisional dissipation of the parton shower energy in the 
QGP. Bands correspond to a range of masses of the propagating system between $m_b$ and $2 m_b$.  
The bottom insert shows the ratio of $R_{AA}$s for radiative + collisional energy loss and
radiative energy loss only. }
\label{RdepRadCol}
\end{figure}

Since a number of CNM and QGP effects can contribute to the $R_{AA}$ of b-jets,
we will study these effects separately to demonstrate how they can be identified and 
constrained from 
the experimental measurements. Let us first concentrate on the redistribution of the 
hard parton energy  
between the vacuum and the medium-induced shower due to soft gluon emission. The physics of 
jet cross section suppression lies in the broader angular distribution of the  medium-induced 
parton shower~\cite{Vitev:2008rz}. Consequently, one expects that the quenching effect 
will be the largest for the 
smallest jet radius $R$ and the effect will disappear ($R_{AA} \rightarrow 1$) for very large radii.  
This is shown in the upper panel of Fig.~\ref{RdepRadCol} for coupling between the jet and the medium 
$g^{\rm med} = 2$ (corresponding to  $\alpha_s^{\rm med} = 0.32$). We concentrate on the region of 
$p_T  > 30$~GeV where hadronization corrections for b-jets are minimal even for small
radii. The jet radius effect of jet quenching is clearly seen by comparing the magnitude of 
the jet suppression for three different radii, $R=0.2$ (red solid line), 
$R=0.4$ (blue dot-dashed line), and $R=0.7$ (light green dashed line). 
The bands correspond to a range of masses for the collimated propagating parent parton 
system  $(m_b,2m_b)$.  The bottom insert in  Fig.~\ref{RdepRadCol} shows the ration
$R_{AA}^{\rm Rad.+Coll.}/ R_{AA}^{\rm Rad.}$ to clarify the significance of the collisional
energy loss for different b-jet radii.

\begin{figure}[t]
\centerline{
\includegraphics[width = 0.95\linewidth]{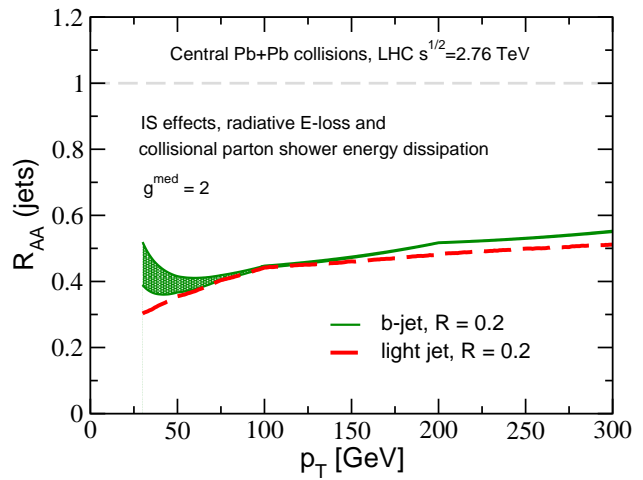}\hskip 0.03\linewidth
}
\caption{The $p_T$-dependent suppressions of both b-jet and light jet cross sections
in central $\sqrt{s_{NN}}=2.76$~TeV  Pb+Pb collisions at the LHC are shown for 
radius $R=0.2$ and the coupling between the jet and the medium $g^{\rm{med}}=2$.
The band for b-jet is the same as in Fig. \ref{RdepRadCol}.}
\label{R02light}
\end{figure}

Note, that above $p_T = 75$~GeV the mass effect disappears even for
$2 m_b = 9$~GeV. This is a direct consequence of the fully coherent energy loss regime.  
For incoherent bremsstrahlung, just like in QED, the mass 
effect never vanishes~\cite{Vitev:2007ve}. Thus, observation of b-jet quenching comparable 
to that of light jets at transverse momenta $p_T > 75$~GeV will constitute direct 
experimental evidence for the dominance of Landau-Pomeranchuk-Migdal type destructive
interference effects in  the medium-induced parton shower formation. 
Below $p_T$ of 75~GeV, there is a distinct trend 
toward reduction of the jet suppression. The reason for this reduction in quenching is two-fold.
On one hand, below $75$~GeV the b-quark mass starts to play a role. On the other hand, the 
b-jet spectra stiffen considerably. Finally, there is a modest $p_T$ dependence 
of $R_{AA}$ up to transverse momenta of 300~GeV.  These features are clearly shown in
Fig.~\ref{R02light}, where a comparison for the nuclear suppression between b-jet and
light jet is presented. The $R_{AA}$ for light-jet production is directly taken from 
previous work~\cite{He:2011pd}. The tiny difference at high $p_T$ is smaller than the 
uncertainty in the treatment of cold nuclear matter effects and collisional energy loss
between these two cases.

In the lower panel of Fig.~\ref{RdepRadCol} we present a similar calculation but include 
the  collisional dissipation of the medium-induced parton shower energy in the QGP.  
This dissipation is 
evaluated as in~\cite{Neufeld:2011yh}, including the interference between the parent parton 
and the radiated gluon, and implemented as thermalization of the soft gluons and transport 
of their energy outside of the jet cone. Clearly, the effect will be most pronounced for 
large radii ($R=0.7$) that contain a significant fraction of the medium-induced parton shower. 
For small radii ($R=0.2$) the effect is negligible. Dissipation of the parton shower energy, 
of course,  still occurs. However, owing to the broad distribution of the medium-induced
shower, which is now verified to ${\cal O}((\alpha_s^{\rm med})^2)$~\cite{Fickinger:2013xwa}, 
this effect is negligible. There simply isn't 
much of the medium-induced shower left inside the jet cone.

\begin{figure}[!b]
\centerline{
\includegraphics[width = 0.95\linewidth]{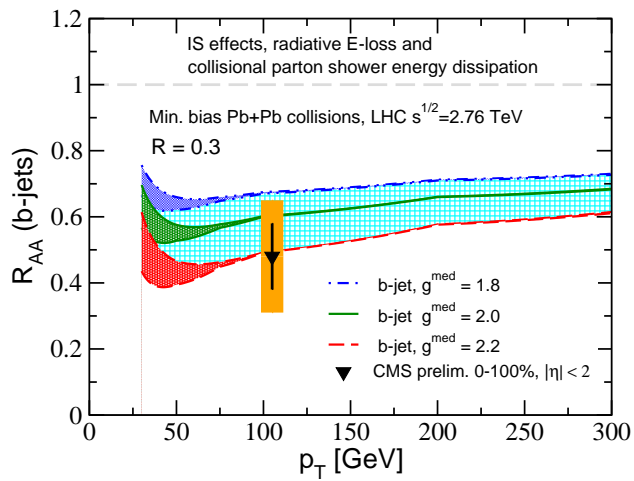}\hskip 0.03\linewidth
}
\caption{The $p_T$-dependent suppression of the b-jet production cross section 
in minimum-bias 
$\sqrt{s_{NN}}=2.76$~TeV  Pb+Pb collisions at the LHC is shown for 3 different couplings
between the jet and the medium, $g^{\rm{med}} = 1.8, 2.0, 2.2$. The pseudeorapidity interval
is $|\eta|<2$ and the jet radius is $R=0.3$.
 Preliminary data is from the CMS collaboration.
 }
\label{R03eta2}
\end{figure}

Next, we turn to the dependence of the nuclear suppression of jet production on the 
coupling between the jet and the strongly-interacting medium. We note that theoretical
predictions for inclusive jet suppression, modification of the di-jet asymmetry distributions,
and $Z^0$-tagged and $\gamma$-tagged jet momentum imbalance distributions with coupling 
constant in the range $g^{\rm med}=2.0 - 2.2$ have compared favourably with the experimental 
findings~\cite{Neufeld:2010fj,He:2011pd}.
We also perform the calculation for an average number of participants 
$N_{\rm part} = 120$ and include isospin effects and cold nuclear matter energy 
loss~\cite{Vitev:2007ve}. Note, 
that at $p_T$ up to $\sim 20 $~GeV in $\sqrt{s_{NN}}=5.08$~TeV p+Pb collisions at the LHC  
CNM effects  appear to be negligible~\cite{ALICE:2012mj}. Above 20~GeV currently there are no 
experimental measurements in p+A reactions at the LHC. The effect of CNM energy 
loss is less than $\sim 15\%$ suppression of b-jet production in minimum bias Pb+Pb collisions
(corresponding roughly to a 7.5\% effect in minimum bias p+A collisions). It also leads to
a weaker $p_T$ dependence of $R_{AA}$. Our simulations are done in the pseudorapidity interval
$ |\eta| < 2 $.  We used three different couplings between the jet and the medium: 
$ g^{\rm med} = 1.8$ (blue dot-dot-dashed line),  $ g^{\rm med} = 2.0$ (green solid line),
 $ g^{\rm med} = 2.2$ (red dashed line).  
The dependence on $g^{\rm med}$ can clearly be identified if experimental
accuracy comparable to that of light jet $R_{AA}$ measurements can be achieved. Shown in 
Fig.~\ref{R03eta2} is a very preliminary CMS measurement of b-jet $R_{AA}$ for 
$p_T > 100$~GeV in the pseudorapidity region $|\eta| < 2$. This preliminary result is for
minimum bias collisions. We observe that there is agreement between the simulations and 
the CMS data point within the large error bars for  $g^{\rm med}$. It is not surprising that 
the suppression of b-jets is similar to the suppression of light quark jets of the 
same transverse momentum. At $p_T  \gtrsim 75 $~GeV the effect of the heavy quark mass,  
$m_b = 4.5$~GeV  to $2 m_b =9$~GeV,  disappears in the coherent energy loss regime. 
On the other hand, the fact that $g\rightarrow b{\bar b}$ system can retain the 
quantum numbers of a gluon implies energy loss considerably larger than that of a 
quark. What would be surprising is if the suppression of b-jet in minimum-bias reaction  
will be equal or larger than the suppression of the light jets in central Pb+Pb 
reactions. 

\begin{figure}[!t]
\centerline{
\includegraphics[width = 0.95\linewidth]{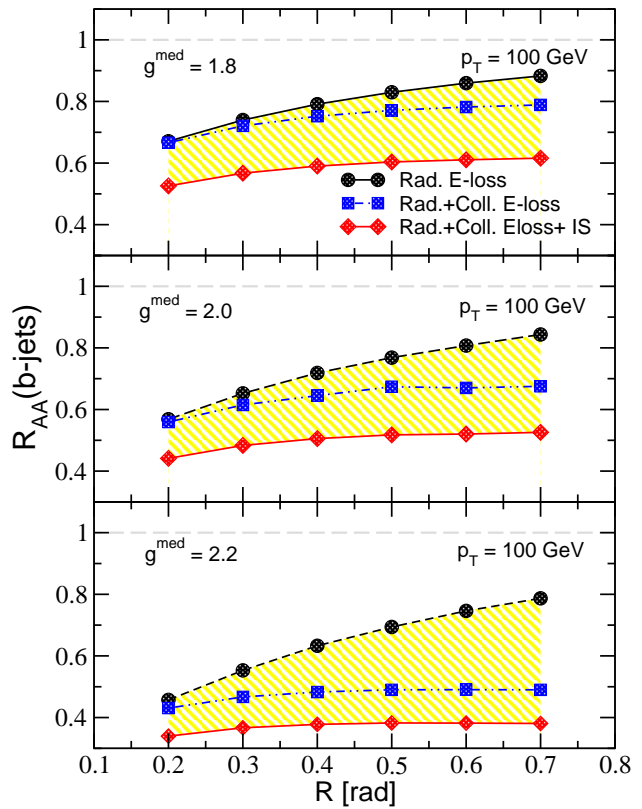}\hskip 0.03\linewidth
}
\caption{$R$ dependence of  $p_T =100$~GeV b-jet suppression as a function of initial-state
and final-state effects. Upper panel is for a coupling between the jet and the medium 
$g^{\rm{med}}=1.8$, middle panel is for $g^{\rm{med}}=2$, and lower panel is for 
$g^{\rm{med}}=2.2$.
}
\label{gDepR}
\end{figure}

Last but not least,  we study the radius dependence and the influence of cold and hot 
nuclear matter effects on the b-jet production in central Pb+Pb collisions.
Our results are shown in Fig.~\ref{gDepR} for $g^{\rm{med}}=1.8$ (upper panel), 
$g^{\rm{med}}=2.0$ (middle panel), and $g^{\rm{med}}=2.2$ (lower panel). For 
definitiveness we fix the jet $p_T = 100$~GeV. The slope of the $R_{AA}(R)$
can help constrain the relative contribution of radiative and collisional energy
loss processes  (redistribution of the 
hard scattered parton energy between a vacuum and a medium-induced parton shower
and  collisional dissipation of the medium-induced parton shower energy in the
QGP).  CNM effects do not change this slope but rather lead to a constant
$\sim 15 - 20\%$ suppression. This creates a degeneracy between the strength of the coupling 
between the jet and the medium and CNM effects. For example, the simulated $R_{AA}$
with $g^{\rm{med}}=2.2$ in the absence of initial state energy loss is practically the same as 
the suppression with $g^{\rm{med}}=2.0$ that includes these inelastic effects. 
This degeneracy cannot be resolved based on Pb+Pb measurements alone. 
b-jet (or even light jet) experimental analysis in p+Pb reactions is needed to 
constrain CNM effects to high transverse momentum at the LHC.

\section{conclusions}
In summary, we presented  theoretical predictions for the nuclear suppression of 
the inclusive b-jet differential cross section in Pb+Pb collisions at 
$\sqrt{s_{NN}}=2.76$~TeV at the LHC. 
Our results combine  b-jet production in p+p, simulated using Pythia~8, 
with the cold nuclear matter and QGP  effects.
The nuclear modification factor $R_{AA}$, as a function of the jet transverse 
momentum $p_T$ and the jet radius parameter $R$, can be compared to the upcoming 
LHC experimental data. We find that, for 
$p_T \gtrsim 75$ GeV, the  nuclear-induced attenuation of b-jets is comparable 
to the one  observed for light jets. The disappearance of the heavy quark mass effects 
at high transverse momentum  is a direct consequence of the fully-coherent final state  
energy loss in the QGP. The detailed radius dependence of $R_{AA}$ at 
fixed $p_T$ provides information for the relative significance of the medium-induced
parton shower formation and the dissipation of its energy in the strongly-interacting 
plasma.  Experimental measurements of b-jet production in heavy ion collisions will,
thus, provide unique new insights into heavy flavor dynamics in dense 
QCD matter.

\section*{Acknowledgments}
We thank Kun Liu, Tao Liu, Yaxian Mao, Feng Wei, and Zhenyu Ye for very helpful 
discussions. This research is supported by the US Department of Energy, Office of 
Science, and in part by the LDRD program at LANL.


\end{document}